\documentclass[twocolumn,showpacs,preprintnumbers,amsmath,amssymb]{revtex4}

\usepackage{graphicx}
\usepackage{dcolumn}
\usepackage{bm}

\begin{document}

\title{Low-frequency $\bm{K^{\pi}=0^{+}}$ modes in deformed neutron-rich nuclei:  
Pairing- and $\bm{\beta}$-vibrational modes of neutron
}

\author{Kenichi Yoshida}
\affiliation{
Department of Physics, Graduate School of Science, Kyoto University, 
Kyoto 606-8502, Japan
}%
\author{Masayuki Yamagami}
\affiliation{
Nishina Center for Accelerator Based Science, 
The Institute of Physical and Chemical Research (RIKEN), 
Wako, Saitama 351-0198, Japan
}%

\date{\today}

\begin{abstract}
Low-frequency $K^{\pi}=0^{+}$ states in deformed neutron-rich nuclei are investigated 
by means of the quasiparticle-random-phase approximation 
based on the Hartree-Fock-Bogoliubov 
formalism in the coordinate space. 
We have obtained the very strongly collective $K^{\pi}=0^{+}$ modes 
not only in neutron-rich Mg isotope but also in Cr and Fe isotopes in $N=40$ region, 
where the onset of nuclear deformation has been discussed. 
It is found that 
the spatially extended structure of neutron quasiparticle wave functions 
around the Fermi level brings about a striking enhancement of the transition strengths. 
It is also found that 
the fluctuation of the pairing field plays an important 
role in generating coherence among two-quasiparticle excitations of neutron.
\end{abstract}

\pacs{21.10.Re; 21.60.Ev; 21.60.Jz}
\maketitle

\section{Introduction}
Physics of nuclei far from $\beta$-stability 
has been one of the main current subjects of nuclear physics. 
Quest for new kinds of collective motion in exotic unstable nuclei 
is one of the most interesting issues in nuclear structure physics 
and has been actively studied both experimentally and theoretically~\cite{COMEX2}. 
Because low-lying collective excitations are sensitive to the shell structure around the 
Fermi level, one can expect unique excitation modes to emerge 
associated with the new spatial structures such as neutron skins 
and novel shell structures which generate new regions of deformation.
 
Breaking of the spherical magic number $N=20$, $28$ and 
striking enhancement of $B(E2;0^{+}_{1} \to 2^{+}_{1})$ in Mg isotopes towards drip line are
under lively discussions in connection with onset of the quadrupole deformation 
and continuum coupling~\cite{mot95,iwa01,yon01,chu05,ele06,bau07,ots01,cau04,ter97,rod02,rei99}. 
Furthermore in heavier mass region, collectivities and breaking of the 
$N=40$ sub-shell have been recently discussed 
for neutron-rich Cr and Fe isotopes~\cite{han99,sor99,sor03,lun07,cau02}.
Although many investigations based on the random phase approximation (RPA) or the Quasiparticle RPA 
(QRPA) including the pairing correlations
~\cite{ham96,ham99,shl03,mat01,hag01,kha02,yam04,mat05,
ter05,ter06,miz07,vre01,paa03,paa04,paa05,cao05,gia03,sar04,per05} 
on multipole responses of neutron-rich nuclei have been done, 
they are largely restricted to spherical systems. 

Recently, low-lying RPA modes in deformed neutron-rich nuclei have been 
investigated by several groups~\cite{nak05,yos05,mut02,ima03,ina06,urk01,hag04a,per07,pen07,nak07,yos08}. 
These calculations, however, do not take into account the pairing correlation, 
or rely on the BCS pairing (except for Ref.~\cite{per07} and our very recent investigation~\cite{yos08}), 
which is inappropriate for describing the pairing correlation 
in drip line nuclei due to the unphysical nucleon gas problem~\cite{dob84}. 

\begin{figure}[t]
\begin{center}
\includegraphics[scale=0.6]{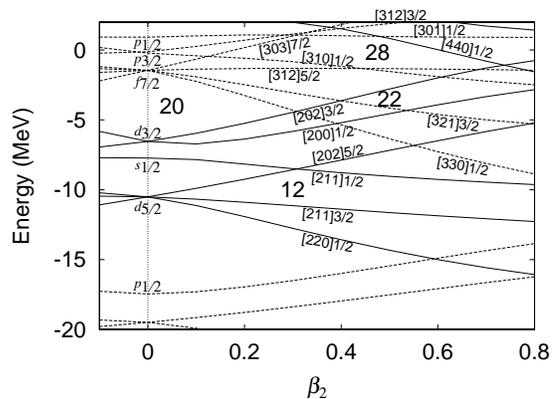}
\caption{Single-particle energies in the deformed WS potential for neutrons 
in $^{40}$Mg, 
plotted as functions of the quadrupole deformation parameter $\beta_{2}$. 
Solid and dotted lines denote positive- and negative-parity levels, respectively. 
Single-particle levels are labeled with the asymptotic quantum numbers $[Nn_{3}\Lambda]\Omega$.}
\label{Fig1}
\end{center}
\end{figure}

In Ref.~\cite{yos06}, we constructed a new computer code that carries out 
the deformed QRPA calculation based on the coordinate-space Hartree-Fock-Bogoliubov 
(HFB) formalism, 
and investigated quadrupole excitations in neutron-rich Mg isotopes close to the drip line 
around $N=28$. 
We obtained low-lying $K^{\pi}=0^{+}$ and $2^{+}$ quadrupole excitation modes in $^{36,38,40}$Mg. 
We found that 
the $K^{\pi}=0^{+}$ mode in $^{40}$Mg was generated by 
coherent superposition of neutron two-quasiparticle (2qp) excitations in 
loosely bound and resonant states. Among them, 
the neutron 2qp excitations of up-sloping oblate levels 
and down-sloping prolate levels play the major role. 
A deformed gap is formed at $N=28$ around $\beta_{2}=0.3$~\cite{rei99} 
due to the crossings between the up-sloping $\nu[303]7/2$ level and 
the down-sloping $\nu[310]1/2$ level, 
and this deformed closed shell 
approximately corresponds to the $(f_{7/2})^{-2}(p_{3/2})^{2}$ configuration
in the spherical shell model representation~\cite{cau04} (see Fig.~\ref{Fig1}).
It was also found that the dynamical pairing correlation, {\it i.e.}, 
fluctuation of the pairing field played a crucial role in 
generating coherence among neutron 2qp excitations. 

The single-particle levels which are $(2j+1)-$fold degenerate in the spherical potential 
are split by the nuclear deformation to 2-fold degenerate levels with the single-particle energy 
to first order of $\beta_{2}$ as~\cite{nil95}
\begin{equation}
\sim \beta_{2} \dfrac{3\Omega^{2}-j(j+1)}{j(j+1)}.
\end{equation} 
In a prolately deformed potential ($\beta_{2}>0$), a level with small value of 
$\Omega$ ($z-$component of the angular momentum $j$) has a negative slope. 
We call this single-particle level a down-sloping prolate level in this paper, 
and refer a positive slope level with large $\Omega$ as a up-sloping oblate level.

Because the deformed shell gap is, in many cases, made by the up-sloping and the 
down-sloping levels~\cite{RS}, we investigate in this paper detailed and systematic properties 
of low-frequency $K^{\pi}=0^{+}$ modes 
in deformed neutron-rich Mg, Cr and Fe isotopes, and discuss generic features of 
the soft $K^{\pi}=0^{+}$ modes uniquely appeared in deformed neutron-rich nuclei.

The paper is organized as follows.
In the next section, we present our model of the deformed HFB and QRPA. 
Numerical results and discussion are given in \S~\ref{results}.
Finally, we summarize the paper in \S~\ref{summary}.

\section{\label{model}Model}
We briefly summarize our approach (see Ref.~\cite{yos06} for details).  
In order to discuss simultaneously effects of nuclear deformation 
and pairing correlations including the continuum, 
we solve the HFB equations~\cite{dob84,bul80,obe03}
\begin{multline}
\begin{pmatrix}
h^{\tau}(\boldsymbol{r}\sigma)-\lambda^{\tau} & \tilde{h}^{\tau}(\boldsymbol{r}\sigma) \\
\tilde{h}^{\tau}(\boldsymbol{r}\sigma) & -(h^{\tau}(\boldsymbol{r}\sigma)-\lambda^{\tau}) \end{pmatrix}
\begin{pmatrix}
\varphi^{\tau}_{1,\alpha}(\boldsymbol{r}\sigma) \\ 
\varphi^{\tau}_{2,\alpha}(\boldsymbol{r}\sigma)
\end{pmatrix}
\\
= E_{\alpha}
\begin{pmatrix}
\varphi^{\tau}_{1,\alpha}(\boldsymbol{r}\sigma) \\ 
\varphi^{\tau}_{2,\alpha}(\boldsymbol{r}\sigma)
\end{pmatrix} \label{eq:HFB1}
\end{multline}
directly in the cylindrical-coordinates 
assuming axial and reflection symmetries. 
Here $\tau=\nu$ (neutron) and $\pi$ (proton), and $\boldsymbol{r}=(\rho,z,\phi)$. 
For the mean-field Hamiltonian $h$, we employ the deformed Woods-Saxon 
potential with the parameters used in Ref.~\cite{yos06} for $^{34}$Mg and 
those in Ref.~\cite{yos05} for neutron-rich Cr and Fe isotopes. 
The pairing field is treated self-consistently by using 
the density-dependent contact interaction~\cite{ber91,ter95}, 
\begin{equation}
v_{pp}(\boldsymbol{r},\boldsymbol{r}^{\prime})=V_{0}\dfrac{1-P_{\sigma}}{2}
\left[ 1-  \dfrac{\varrho^{\mathrm{IS}}(\boldsymbol{r})}{\varrho_{0}} \right]
\delta(\boldsymbol{r}-\boldsymbol{r}^{\prime}). \label{eq:res_pp}
\end{equation}
with $\varrho_{0}=0.16$ fm$^{-3}$. 
Here $\varrho^{\mathrm{IS}}(\boldsymbol{r})$ denotes the isoscalar density and 
$P_{\sigma}$ the spin exchange operator. 
Because the time-reversal symmetry and the reflection symmetry 
with respect to the $x-y$ plane are assumed, 
we have only to solve for positive $\Omega$ and positive $z$. 
We use the lattice mesh size $\Delta\rho=\Delta z=0.8$ fm 
and the box boundary condition at  
$\rho_{\mathrm{max}}=10.0$ fm and $z_{\mathrm{max}}=12.8$ fm. 
The quasiparticle energy is cut off at 50 MeV and 
the quasiparticle states up to $\Omega^{\pi}=15/2^{\pm}$ are included. 

Using the quasiparticle basis obtained by solving the HFB equation (\ref{eq:HFB1}),  
we solve the QRPA equation in the matrix formulation~\cite{row70} 
\begin{equation}
\sum_{\gamma \delta}
\begin{pmatrix}
A_{\alpha \beta \gamma \delta} & B_{\alpha \beta \gamma \delta} \\
B_{\alpha \beta \gamma \delta} & A_{\alpha \beta \gamma \delta}
\end{pmatrix}
\begin{pmatrix}
f_{\gamma \delta}^{\lambda} \\ g_{\gamma \delta}^{\lambda}
\end{pmatrix}
=\hbar \omega_{\lambda}
\begin{pmatrix}
1 & 0 \\ 0 & -1
\end{pmatrix}
\begin{pmatrix}
f_{\alpha \beta}^{\lambda} \\ g_{\alpha \beta}^{\lambda}
\end{pmatrix} \label{eq:AB1}.
\end{equation}
The residual interaction in the particle-particle (p-p) channel 
appearing in the QRPA matrices $A$ and $B$ is 
the density-dependent contact interaction (\ref{eq:res_pp}).
On the other hand, 
for the residual interaction in the particle-hole (p-h) channel, 
we use the Skyrme-type interaction 
\begin{equation}
v_{ph}(\boldsymbol{r},\boldsymbol{r}^{\prime})=
\left[ t_{0}(1+x_{0}P_{\sigma})+\dfrac{t_{3}}{6}(1+x_{3}P_{\sigma})
\varrho^{\mathrm{IS}}(\boldsymbol{r}) \right]
\delta(\boldsymbol{r}-\boldsymbol{r}^{\prime}), \label{eq:res_ph}
\end{equation}
with $t_{0}=-1100$ MeV$\cdot$fm$^{3}$, $t_{3}=16000$ MeV$\cdot$fm$^{6}, 
x_{0}=0.5$, and $x_{3}=1.0$~\cite{shl75}.
Because the deformed Wood-Saxon potential is used for the mean field, 
we renormalize the residual interaction in the p-h channel 
by multiplying a factor $f_{ph}$
to get the spurious $K^{\pi}=1^{+}$  mode
(representing the rotational mode) at zero energy
($v_{ph} \rightarrow f_{ph}\cdot v_{ph}$).  
We cut the 2qp space at $E_{\alpha}+E_{\beta} \leq 30$ MeV 
due to the excessively demanding computer memory for the model space 
consistent with that adopted in the HFB calculation. 
Accordingly, we need another factor $f_{pp}$
for the p-p channel. 
We determine this factor such that the spurious $K^{\pi}=0^{+}$ mode 
associated with the particle number fluctuation appears at zero energy
($v_{pp} \rightarrow f_{pp}\cdot v_{pp}$). 
The dimension of the QRPA equation~(\ref{eq:AB1}) is 2512 for $^{34}$Mg and 
3804 for $^{72}$Fe. 
Within this model space, the energy-weighted sum rule for the quadrupole 
$K^{\pi}=0^{+}$ excitation is satisfied $92.9 \%$ for $^{34}$Mg 
and $97.0 \%$ for $^{72}$Fe. 

\section{\label{results}Results and Discussion}
Let us now discuss the low-frequency $K^{\pi}=0^{+}$ modes in 
$^{34}$Mg in \S~\ref{Mg} and in neutron-rich Cr and Fe isotopes 
with $N \simeq 40$ in \S~\ref{CrFe}.

\subsection{\label{Mg}${}^{34}Mg$}
\begin{figure}[t]
\begin{center}
\includegraphics[scale=0.6]{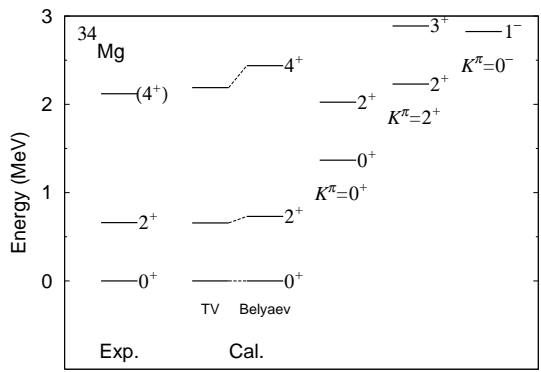}		
\caption{Excitation energy spectrum in $^{34}$Mg obtained by the QRPA calculation 
using $\beta_{2}=0.4$ and $V_{0}=-400$ MeV$\cdot$fm$^{3}$ 
and available experimental data~\cite{yon01}.  
}
\label{34Mg_level}
\end{center}
\end{figure}

Figure~\ref{Fig1} shows the single-particle energy diagram 
for the WS potential as a function of deformation parameter $\beta_{2}$. 
At around $\beta_{2}=0.5$, the deformed shell gap is formed at $N=22$ 
due to the crossings between the up-sloping $\nu[202]3/2$ and the 
down-sloping $\nu[330]1/2$ and $\nu[321]3/2$ levels. 
In Fig.~\ref{34Mg_level}, we show the low-lying excitation spectrum below 3 MeV in $^{34}$Mg. 
Here excitation energies are evaluated by~\cite{EG} 
\begin{equation}
E(I,K)=\hbar \omega_{\mathrm{RPA}}+\frac{\hbar^{2}}{2\mathcal{J}_{\mathrm{TV}}}(I(I+1)-K^{2}),
\end{equation}
with the vibrational frequencies, $\omega_{\mathrm{RPA}}$, and 
the Thouless-Valatin moment of inertia, $\mathcal{J}_{\mathrm{TV}}$,
calculated by the QRPA. 
The pairing strength $V_{0}$ for $^{34}$Mg is determined in order to reproduce the experimental 
excitation energy of $E(2_{1}^{+})=660$ keV~\cite{yon01}. 
In this figure, we also show the excitation energies evaluated by using the 
Belyaev moment of inertia, $\mathcal{J}_{\mathrm{Belyaev}}$, 
where the residual interaction is turned off. 
We found that $\mathcal{J}_{\mathrm{TV}}$ becomes about $11 \%$ larger than 
$\mathcal{J}_{\mathrm{Belyaev}}$ due to the time-odd component in the residual interactions 
of (\ref{eq:res_pp}) and (\ref{eq:res_ph}). 
The $K^{\pi}=2^{+}$ mode at 2.01 MeV is dominantly generated by the proton p-h excitations 
as that in $^{36,38,40}$Mg~\cite{yos06}; these are proton dominated $\gamma-$vibrational modes.
The negative parity $K^{\pi}=0^{-}$ state at 2.61 MeV 
is mainly generated by the lowest neutron 2qp excitation of 
$\nu[202]3/2 \otimes \nu[321]3/2$ with the excitation energy of 2.74 MeV. 
Its collectivity is thus rather weak.

\begin{figure}[t]
  \begin{center}
    \begin{tabular}{ccc}
	\includegraphics[height=4.5cm]{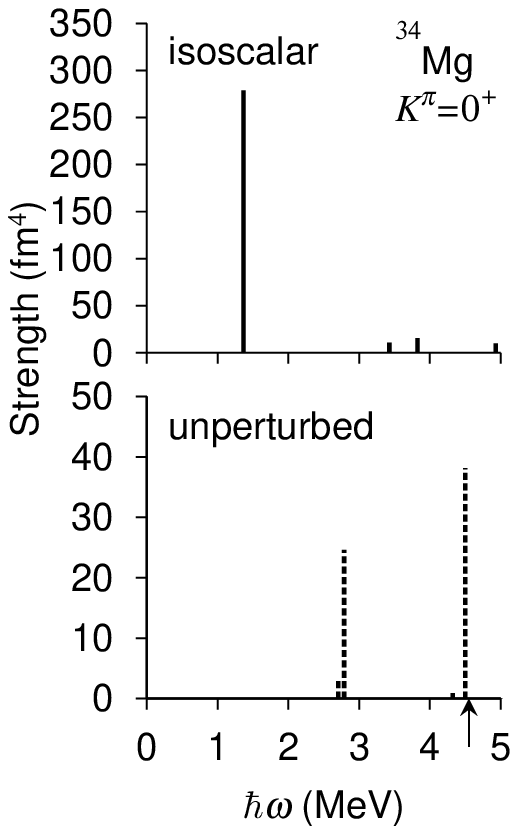}
	\includegraphics[height=4.5cm]{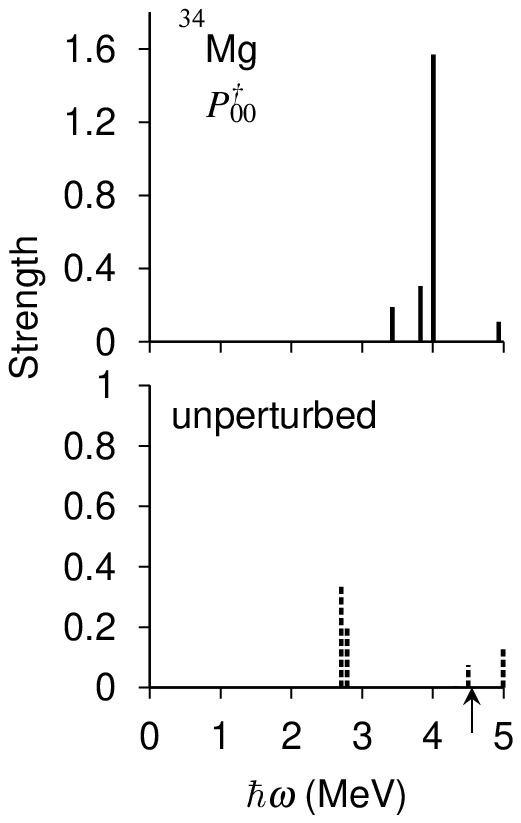}
	\includegraphics[height=4.5cm]{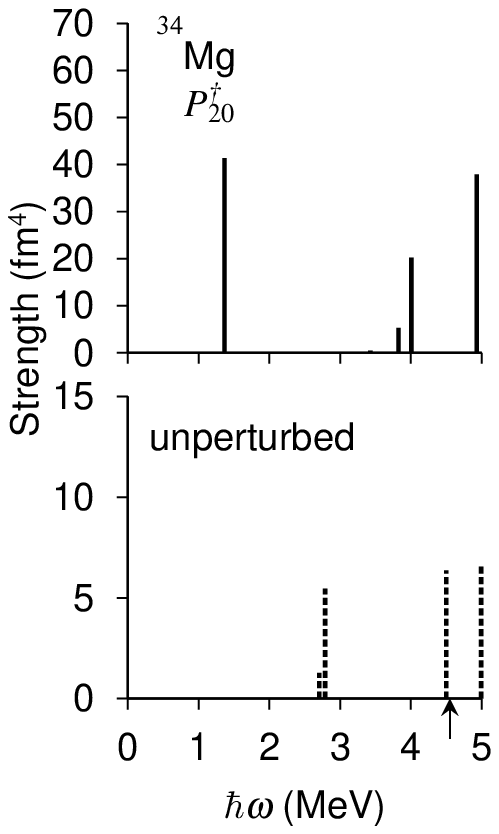}
    \end{tabular}
\caption{QRPA strength distributions for the  $K^{\pi}=0^{+}$ 
isoscalar quadrupole p-h excitations (left) and 
the monopole- and quadrupole-pair excitations (middle and right) in $^{34}$Mg, 
calculated using $f_{ph}=0.74$ and $f_{pp}=1.37$. 
For comparison, unperturbed 2qp transition strengths are shown 
in the lower panels. The arrow indicates the neutron threshold energy 
$E_{\mathrm{th}}=4.55$ MeV for 1qp continuum.}
\label{34Mg_strength}
\end{center}
\end{figure}

The left panel of Fig.~\ref{34Mg_strength} shows 
the $K^{\pi}=0^{+}$ isoscalar quadrupole transition strengths, 
and the unperturbed 2qp transition strengths are shown in the lower panel.
The intrinsic transition strength to the lowest $K^{\pi}=0^{+}$ excited state 
is about 43 Weisskopf units (1 W.u. $\simeq 6.5$ fm$^{4}$ for $^{34}$Mg). 
This state is constructed by the coherent superposition of 
2qp excitations of the up-sloping $(\nu[202]3/2)^{2}$ and 
the down-sloping $(\nu[330]1/2)^{2}$ and the $(\nu[321]3/2)^{2}$ levels, 
together with the high-lying $2\hbar \omega$ excitation of protons.

\begin{table*}[t]
\caption{QRPA amplitudes for the $K^{\pi}=0^{+}$ state at 1.37 MeV in $^{34}$Mg.
This mode has the proton strength $B(E2)=15.9 e^{2}$fm$^{4}$, 
the neutron strength $B(Q^{\nu}2)=162 $fm$^{4}$, 
and the isoscalar strength $B(Q^{\mathrm{IS}}2)=279 $fm$^{4}$, 
and the sum of backward-going amplitudes $\sum|g_{\alpha\beta}|^{2}=0.141$. 
The single-quasiparticle levels are labeled with 
the asymptotic quantum numbers $[Nn_{3}\Lambda]\Omega$. 
Only components with $f_{\alpha\beta}^{2}-g_{\alpha\beta}^{2} > 0.01$ are listed. 
}
\label{34Mg_0+}
\begin{center} 
\begin{tabular}{cccccccccccc}
\hline \hline
 &  &  & $E_{\alpha}+E_{\beta}$ &  & & 
$Q_{20,\alpha\beta}^{\mathrm{(uv)}}$ & $M_{20,\alpha\beta}^{\mathrm{(uv)}}$ &
$P_{20,\alpha\beta}^{\mathrm{(uu)}}$ & $M_{20,\alpha\beta}^{\mathrm{(add)}}$ &
$P_{00,\alpha\beta}^{\mathrm{(uu)}}$ & $M_{00,\alpha\beta}^{\mathrm{(add)}}$ \\
 & $\alpha$ & $\beta$ & (MeV) & $f_{\alpha \beta}$ & $g_{\alpha\beta}$ & (fm$^{2}$) & (fm$^{2}$) &
 (fm$^{2}$) & (fm$^{2}$) &  &  \\ \hline
(a) & $\nu[202]3/2$ & $\nu[202]3/2$ & 2.70 & $-0.615$ & $-0.230$  &  $-1.814$ & 1.533 & $-1.128$ & 0.864 & 0.579 & $-0.453$ \\ 
(b) & $\nu[321]3/2$ & $\nu[321]3/2$ & 2.79 & 0.688 & 0.052 & 4.962 & 3.673 & 2.335 & 1.745 & 0.443 & 0.334 \\
(c) & $\nu[330]1/2$ & $\nu[330]1/2$ & 4.50 & 0.371 & 0.062 & 6.177 & 2.676 & 2.524 & 1.329 & 0.278 & 0.146 \\
\hline \hline
\end{tabular}
\end{center} 
\end{table*}

Generation mechanism of the soft $K^{\pi}=0^{+}$ mode in deformed neutron-rich nuclei is 
understood essentially by the schematic two-level model in Ref.~\cite{BM2}.
They consider the case where only two $\lambda \bar{\lambda}$ components 
are present in the wave functions both of the ground $0^{+}_{\mathrm{gs}}$ and 
of the excited $0^{+}_{2}$ states;
\begin{subequations}
\begin{align}
|0^{+}_{\mathrm{gs}}\rangle &= \dfrac{a}{\sqrt{a^2+b^2}}|\lambda_{1}\bar{\lambda}_{1}\rangle + 
\dfrac{b}{\sqrt{a^2+b^2}}|\lambda_{2}\bar{\lambda}_{2}\rangle \\
|0^{+}_{2}\rangle &= -\dfrac{b}{\sqrt{a^2+b^2}}|\lambda_{1}\bar{\lambda}_{1}\rangle + 
\dfrac{a}{\sqrt{a^2+b^2}}|\lambda_{2}\bar{\lambda}_{2}\rangle.
\end{align}
\end{subequations}
The transition matrix element for the quadrupole operator is then 
\begin{equation}
\langle 0^{+}_{\mathrm{gs}}|\hat{Q}_{20}|0^{+}_{2}\rangle =
\dfrac{2ab}{a^{2}+b^{2}}[ \langle \lambda_{1}|\hat{Q}_{20}|\lambda_{1}\rangle 
- \langle \lambda_{2}|\hat{Q}_{20}|\lambda_{2}\rangle ]
\end{equation}
and it is proportional to the difference in the 
quadrupole moments of the individual orbitals composing the $0^{+}$ states.
In the case that the quadrupole moments of the orbitals have opposite signs to each other, 
this matrix element becomes large. 
This situation is realized in the level crossing region, where the up- and down-sloping 
orbitals exist. 
As the number of components increases in the QRPA calculations, 
the wave function becomes more complicated, 
and the transition matrix element reads (see Appendix B)
\begin{equation}
\langle K^{\pi}=0^{+} | \hat{Q}_{20} | 0 \rangle =
\sum_{\alpha\beta} Q_{20,\alpha\beta}^{\mathrm{(uv)}}(f_{\alpha\beta}+g_{\alpha\beta})
\equiv \sum_{\alpha\beta}M_{20,\alpha\beta}^{\mathrm{(uv)}}. \label{matrix_element}
\end{equation}

In the QRPA employing the separable interaction, 
for the major component of the QRPA eigenmode associated with the pairing fluctuation, 
the phase of the QRPA amplitudes, 
$f_{\alpha \beta}$ and $g_{\alpha \beta}$, are
opposite between particle-like levels and hole-like levels
(due to the $(u^2-v^2)$ factor; see {\it Appendix J} of Ref.~\cite{bri05}).
In the level crossing region, the particle-like level is the up-sloping oblate level, 
and the hole-like level is the down-sloping prolate level. 
Because the transition matrix elements for the QRPA eigenmodes are 
determined by the sum of products of the QRPA amplitudes and the 
individual 2qp transition matrix elements in Eq.~(\ref{matrix_element}), 
they contribute coherently for the quadrupole transition.

In the present case, 
the main components generating the low-lying $K^{\pi}=0^{+}$ state are 
$[202]3/2\otimes\overline{[202]3/2}$, $[321]3/2\otimes \overline{[321]3/2}$ and 
$[330]1/2\otimes\overline{[330]1/2}$, 
which are the up-sloping and down-sloping levels with opposite quadrupole moments 
and the QRPA amplitudes have opposite signs as shown in Table~\ref{34Mg_0+}. 
Therefore, the strength for the quadrupole transition becomes enhanced. 
These three 2qp excitations contribute about 50~\% of the total transition strength. 
Furthermore, many other 2qp excitations coherently participate to generate the lowest $0^{+}$ state, 
which brings about further enhancement of the transition strength. 

\begin{figure}[t]
\begin{center}
\includegraphics[scale=0.47]{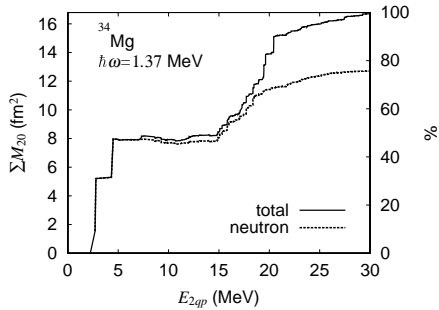}
\caption{Sum of the transition matrix element $\sum_{\alpha\beta} M_{20,\alpha\beta}^{(\mathrm{uv})}$ 
for the lowest $K^{\pi}=0^{+}$ state at 1.37 MeV. 
Solid and dotted lines denote the isoscalar and neutron transitions, respectively.}
\label{34Mg_analysis}
\end{center}
\end{figure}

Figure~\ref{34Mg_analysis} shows the partial sum of the transition matrix element 
for the $K^{\pi}=0^{+}$ mode
$\sum_{ E_{\alpha}+E_{\beta} \leqq E_{\mathrm{2qp}}}M_{20,\alpha\beta}^{(\mathrm{uv})}$
for the isoscalar and neutron excitations. 
We see that many 2qp excitations construct this state with coherence; 
each 2qp transition matrix element ($M_{20,\alpha\beta}^{\mathrm{(uv)}}$) has a same sign.
If the QRPA mode were a destructive state or a single 2qp excitation, 
the sum of the transition matrix element becomes zero in total or shows step function. 

From this figure, we see another interesting feature: 
Soft $K^{\pi}=0^{+}$ mode in $^{34}$Mg is generated not only by neutron 2qp excitations 
around the Fermi level with the excitation energy around 5 MeV (three 2qp excitations discussed above) 
but also proton excitations around 20 MeV. 
These proton excitations are $2\hbar \omega$ excitations of giant resonance; 
$\pi[211]3/2 \to \pi[431]3/2$ (19.5 MeV), $\pi[101]3/2 \to \pi[321]3/2$ (17.3 MeV), 
$\pi[110]1/2 \to \pi[330]1/2$ (19.6 MeV) and $\pi[220]1/2 \to \pi[440]1/2$ (20.1 MeV).  
Due to this coupling with giant resonance of proton, the proton transition 
strength $B(E2)$ is larger than that in $^{40}$Mg; $B(E2)=15.9 e^{2}$fm$^{4}$ 
in $^{34}$Mg and $3.4 e^{2}$fm$^{4}$ in $^{40}$Mg~\cite{yos06}. 
Hindrance of the proton excitation in drip-line nuclei is due to the 
extreme spatial extension of neutron wave functions and 
decoupling with proton wave functions~\cite{yos06}.

The discussion for enhancement of the quadrupole transition strength to 
the low-lying $K^{\pi}=0^{+}$ mode 
is also applicable to the pair transition strength.
In the middle and right panels of Fig.~\ref{34Mg_strength}, 
we show two kinds of strength for neutron pair transitions; 
one is the monopole pair transition 
and another the quadrupole pair transition, 
\begin{subequations}
\begin{align}
\hat{P}^{\dagger}_{00}&=\int d\boldsymbol{r}
\hat{\psi}^{\dagger}(\boldsymbol{r}\uparrow)\hat{\psi}^{\dagger}(\boldsymbol{r}\downarrow), \\
\hat{P}^{\dagger}_{20}&=\int d\boldsymbol{r} r^{2}Y_{20}(\hat{r})
\hat{\psi}^{\dagger}(\boldsymbol{r}\uparrow)\hat{\psi}^{\dagger}(\boldsymbol{r}\downarrow). 
\end{align} \label{eq:pair_transition}
\end{subequations}
For the quadrupole pair transitions, we see a prominent peak 
at 1.37 MeV, whereas no peak is seen at this energy 
for the monopole pair transitions.
The origin of this contrasting behavior is understood as follows: 
The quadrupole (monopole) pairing matrix elements for individual 2qp 
excitations are in opposite phase (in phase) between the down-sloping 
prolate levels and the up-sloping oblate levels. 
As the transition matrix elements for the QRPA eigenmodes are 
determined by the similar way for the p-h transition matrix elements (See Eq.~(B4)), 
they contribute coherently and destructively for the quadrupole and monopole pair transitions, 
respectively.
As a consequence, the quadrupole-pair transition strength 
to the lowest $K^{\pi}=0^{+}$ collective excited state is greatly enhanced, 
whereas the monopole-pair transition strength to this state almost vanishes. 

Therefore, 
in a deformed system where the up- and down-sloping orbitals exist near the Fermi level and 
the pairing fluctuation becomes important, 
we can expect the emergence of the low-lying mode which has extremely enhanced strengths 
both for the quadrupole p-h and for the quadrupole p-p (pair) transition. 

\begin{figure}[t]
\begin{center}
\includegraphics[scale=0.52]{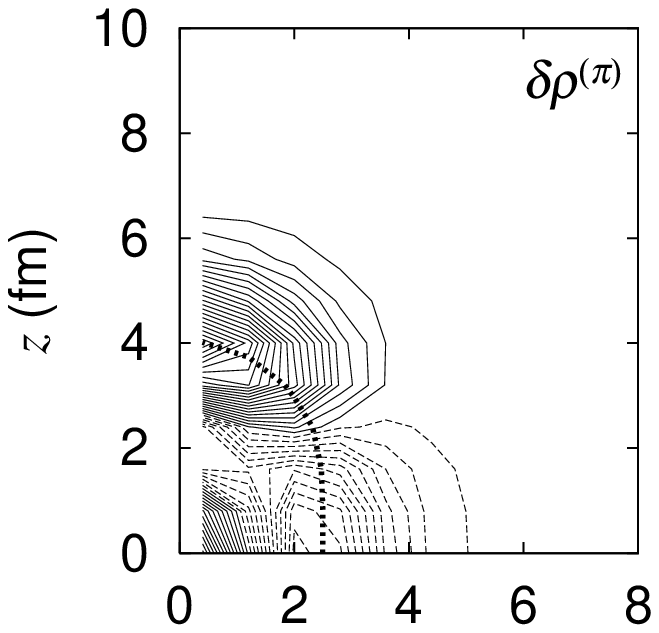}
\includegraphics[scale=0.52]{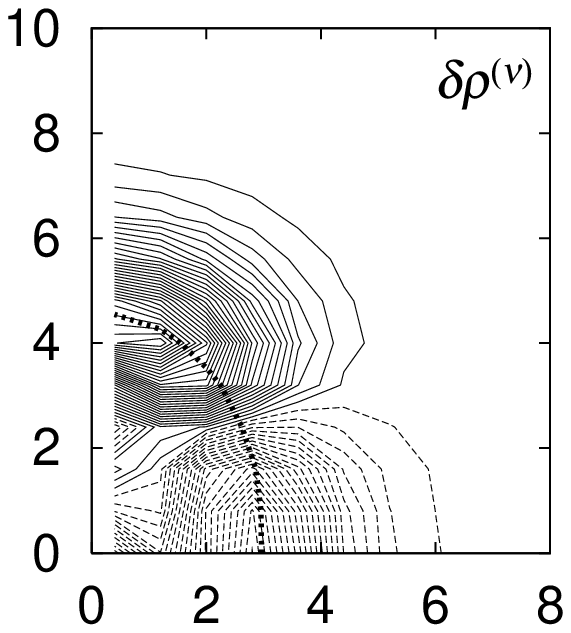} \\
\includegraphics[scale=0.52]{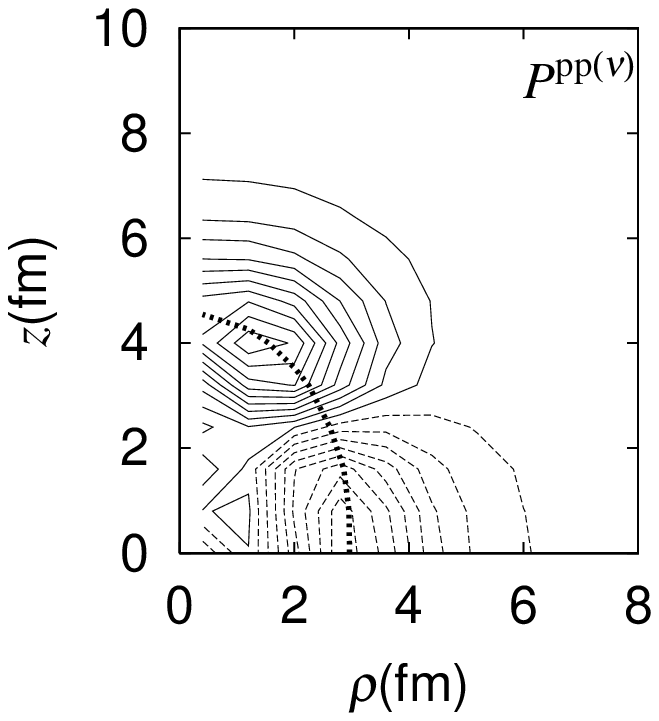}
\includegraphics[scale=0.52]{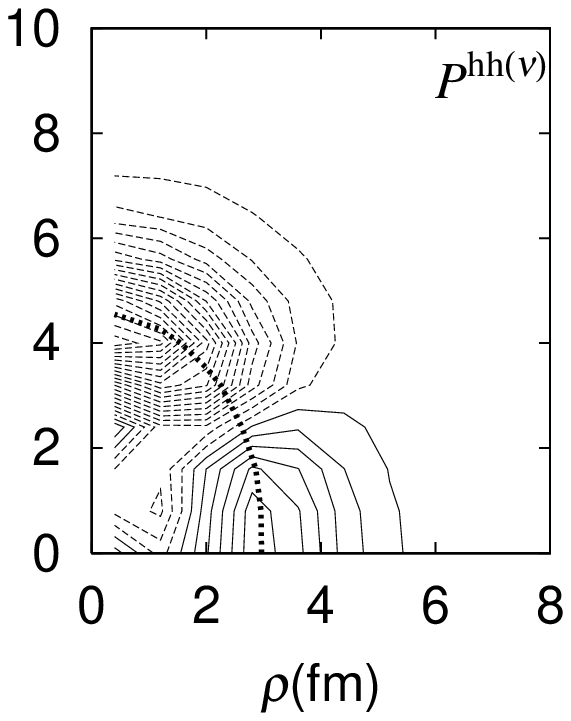}
\caption{Transition densities to the $K^{\pi}=0^{+}$ state at 1.37 MeV. 
The particle-hole transition densities of protons and neutrons (upper panels), 
the particle-particle and hole-hole transition densities of neutrons (lower panels). 
Solid and dashed lines indicate positive 
and negative transition densities, and 
the contour lines are plotted at intervals of $3 \times 10^{-4}$ fm$^{-3}$. 
The thick dotted lines denote the proton and neutron half densities, 
0.041 and 0.049 fm$^{-3}$, respectively.
}
\label{transition_density}
\end{center}
\end{figure}

The low-lying $K^{\pi}=0^{+}$ modes possibly emerge both in stable and in unstable deformed nuclei, 
and the unique feature in neutron-rich nuclei is that they are 
mainly generated by neutrons, whose wave functions have spatially extended 
structure due to the shallow Fermi level.
In order to see the unique spatial structure, 
we show in Fig.~\ref{transition_density} three kinds of transition densities;
\begin{subequations}
\begin{align}
\delta \varrho^{(\tau)} (\rho,z)&=\langle \lambda |\sum_{\sigma}
\psi^{\dagger}_{\tau}(\rho,z,\sigma)\psi_{\tau}(\rho,z,\sigma)
|0\rangle, \\
P^{\mathrm{pp} (\nu)}(\rho,z)&=\langle \lambda| 
\psi^{\dagger}_{\nu}(\rho,z,\uparrow)\psi^{\dagger}_{\nu}(\rho,z,\downarrow)|0\rangle, \\
P^{\mathrm{hh} (\nu)}(\rho,z)&=\langle \lambda|
\psi_{\nu}(\rho,z,\downarrow)\psi_{\nu}(\rho,z,\uparrow)|0\rangle,
\end{align} 
\end{subequations}
where they are called the p-h transition density, the p-p (particle-pair) 
transition density, and the h-h (hole-pair) transition density, respectively~\cite{mat05}. 
The upper panels show a typical shape vibration of protons and neutrons 
along the symmetry axis ($z-$axis). 
We can clearly see the spatial extension of the neutron wave functions; 
the excitation takes place in the skin region around 7--8 fm. 
This spatial extension of the neutron wave functions brings about the enhancement 
of the transition strength. 
Another unique feature of the low-lying $K^{\pi}=0^{+}$ mode 
is that the particle-pair and hole-pair 
transition densities have an appreciable amplitude and are almost comparable to 
the p-h transition density. 

\subsection{\label{CrFe}Cr and Fe isotopes around $N=40$}
\begin{figure}[t]
\begin{center}
\includegraphics[scale=0.6]{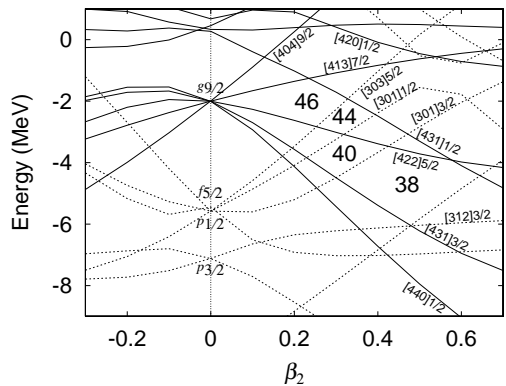}
\caption{Same as Fig.~\ref{Fig1} but in $^{64}$Cr.}
\label{Fig6}
\end{center}
\end{figure}

\begin{figure*}[t]
  \begin{center}
   \begin{tabular}{c}
      \includegraphics[height=5.2cm]{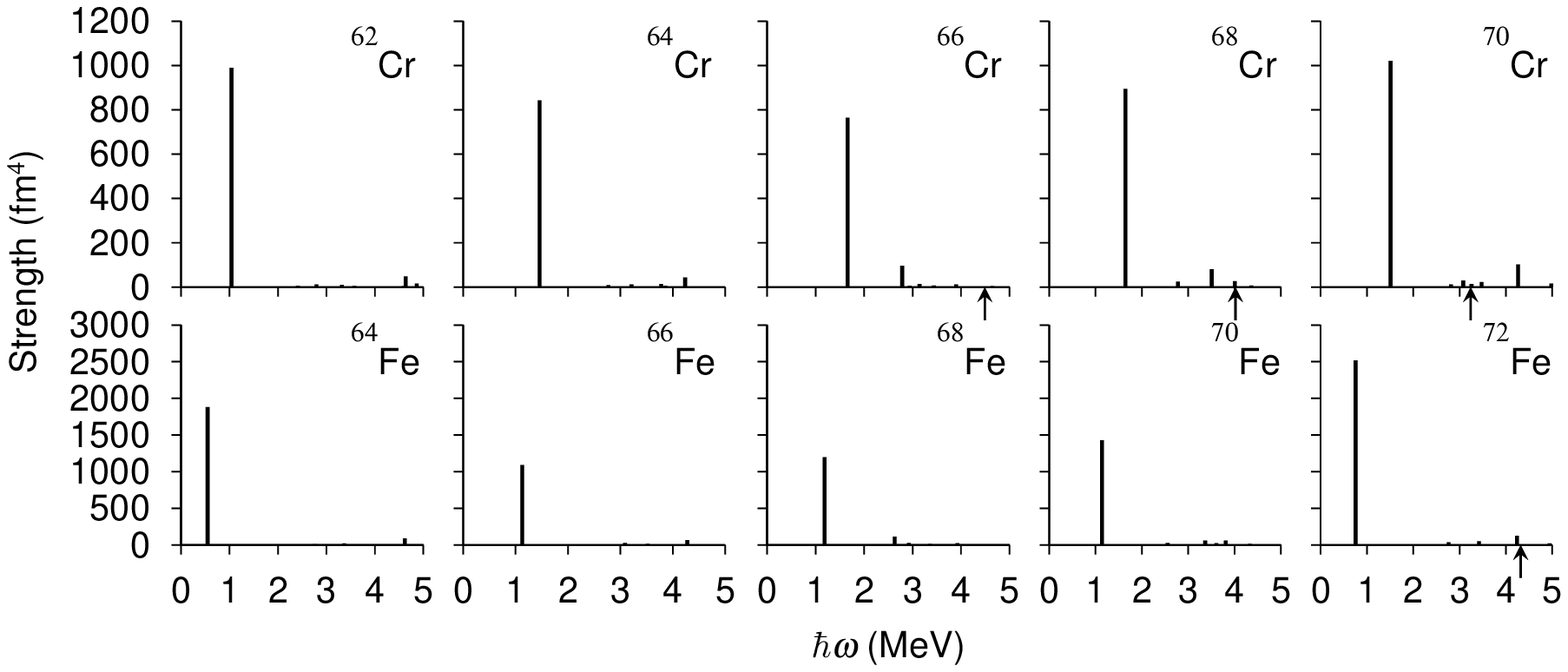}\\
	\includegraphics[height=5.2cm]{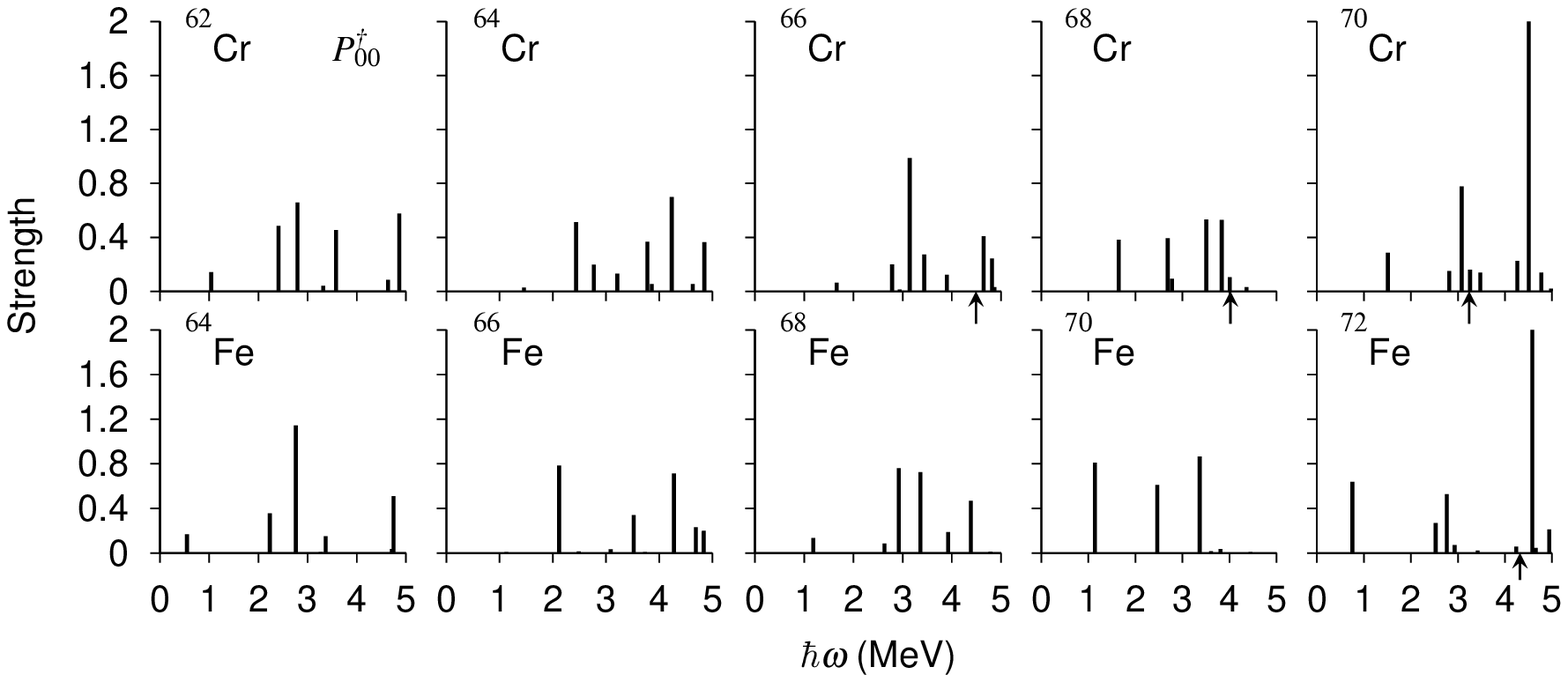}\\
	\includegraphics[height=5.2cm]{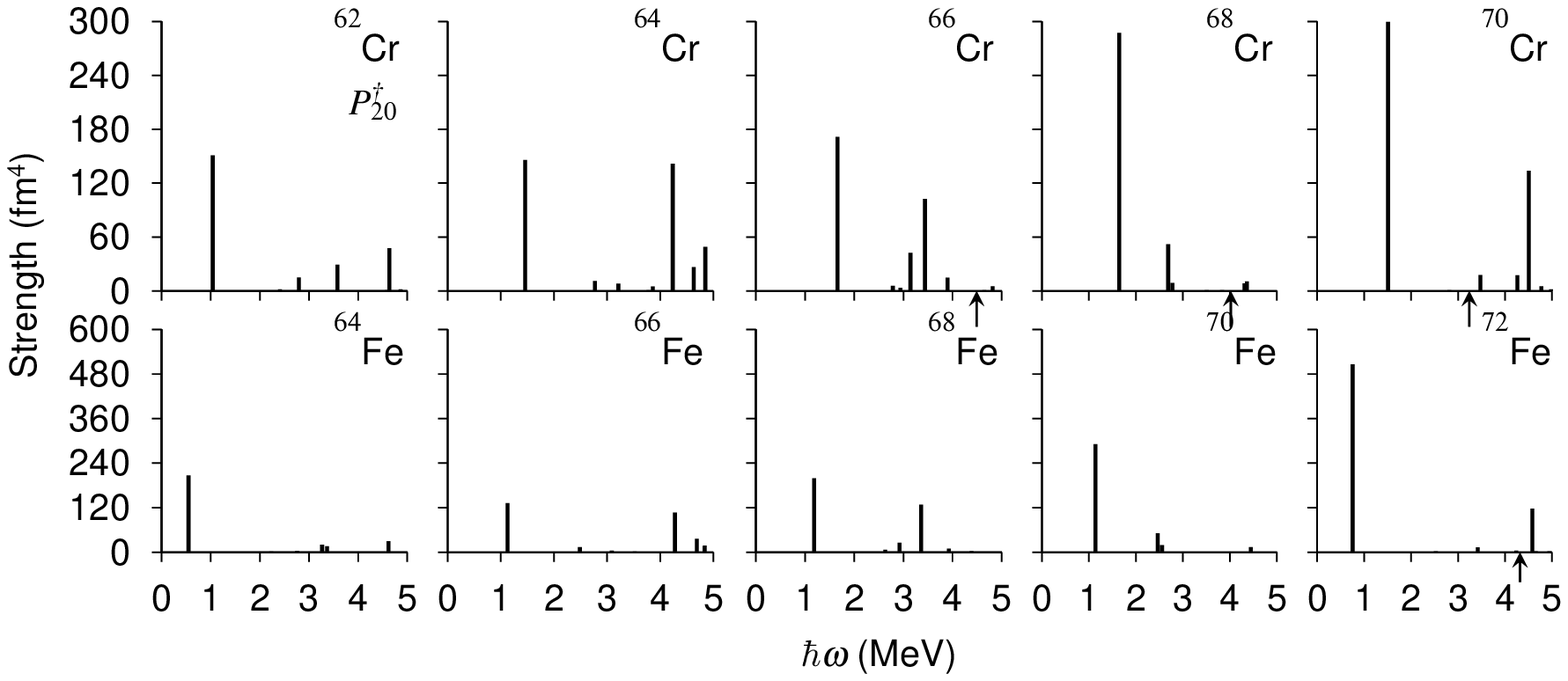}
   \end{tabular}		
    \caption{Isoscalar quadrupole (top), the monopole-pair (middle), 
and the quadrupole-pair (bottom) transition strengths for the $K^{\pi}=0^{+}$ excitations in 
$^{62-70}$Cr and $^{64-72}$Fe obtained by the QRPA calculations using $\beta_{2}=0.3$ and 
$V_{0}=-420$ MeV$\cdot$fm$^{3}$. The self-consistent factors are $f_{ph}$=0.72--0.74 and 
$f_{pp}=1.43$ for Cr isotopes, and $f_{ph}$=0.74--0.79 and $f_{pp}$=1.39--1.42 for Fe isotopes. 
The arrows denote the neutron threshold energies.
}
\label{Cr_Fe_strength}
\end{center}
\end{figure*}

The deformation region near the shell closure gives a favorable situation for emergence of 
the soft $K^{\pi}=0^{+}$ modes discussed in the previous subsection 
because many orbitals with opposite quadrupole moments are crossing around 
the Fermi level. 
Since the low-lying $K^{\pi}=0^{+}$ mode is quite sensitive to the pairing correlation 
and deformation, 
we investigate in the present subsection the low-frequency $K^{\pi}=0^{+}$ excitations 
in neutron-rich Cr and Fe isotopes in $N=40$ region, 
and show generic features of the soft $K^{\pi}=0^{+}$ modes in deformed neutron-rich nuclei.

In a prolately deformation region around $\beta_{2}\simeq 0.3$, 
deformed shell gaps associated with different $\beta_2$ values 
appear at various neutron numbers due to the crossings between 
the down-sloping $\nu[440]1/2$, $\nu[431]3/2$ and $\nu[431]1/2$ levels 
and the up-sloping $\nu[301]3/2$, $\nu[301]1/2$ and $\nu[303]5/2$ 
levels (see Fig.~\ref{Fig6}). 

According to the Skyrme HFB calculations~\cite{sto03,sto05}, 
the ground state of the neutron-rich Cr and Fe isotopes with $N \simeq 40$ has a 
finite deformation around $\beta_{2}=0.3$.

The top panels in Fig.~\ref{Cr_Fe_strength} show the isoscalar quadrupole transition strengths 
for the $K^{\pi}=0^{+}$ excitations. 
We can see a prominent peak at around 1 MeV in all nuclei under consideration. 
All low-lying states obtained here have extremely enhanced transition strengths
(note that 1 W. u. is 14.6--17.1 fm$^{4}$ for $^{62-70}$Cr and 
15.2--17.8 fm$^{4}$ for $^{64-72}$Fe).
These excitation modes are generated by coherent superposition of neutron 2qp 
excitations among the up- and the down-sloping levels as in $^{34}$Mg.

Furthermore, we can see an interesting feature from this systematic calculation; 
the transition strengths become enhanced at $N=38$ and 46, 
and are symmetric at $N=42$. 
This feature can be understood by the role of the $\Omega^{\pi}=1/2^{+}$ levels. 
In $^{62}$Cr and $^{64}$Fe, 
2qp excitations of the up-sloping $(\nu[301]3/2)^{2}$ and the down-sloping $(\nu[440]1/2)^{2}$
have largest contributions to the lowest $K^{\pi}=0^{+}$ state. 
The transition strength of the $(\nu[440]1/2)^{2}$ excitation is 86 fm$^{4}$ (5.9 W.u.) in $^{62}$Cr 
and 64 fm$^{4}$ (4.2 W.u.) in $^{64}$Fe, 
whereas that of the $(\nu[301]3/2)^{2}$ is 9.6 fm$^{4}$ and 8.6 fm$^{4}$, respectively. 
This indicates the $\nu[440]1/2$ level has a spatially extended structure 
(in $^{62}$Cr, the root-mean-square radius of the $\nu[440]1/2$ level is 5.1 fm, whereas 
the total neutron r.m.s. radius is 4.2 fm). 
As the neutron number increases, 
the transition strength decreases as a consequence of the decreasing 
contribution of the $(\nu[440]1/2)^{2}$ excitation.   

As the neutron number approaches $N=46$, the $\nu[431]1/2$ level becomes 
close to the Fermi surface (see Fig.~\ref{Fig6}). 
In $^{70}$Cr and $^{72}$Fe, 
2qp excitations of the up-sloping $(\nu[303]5/2)^{2}$ and the down-sloping $(\nu[431]1/2)^{2}$
have largest contributions to the lowest $K^{\pi}=0^{+}$ state. 
The transition strength of the $(\nu[431]1/2)^{2}$ excitation is 115 fm$^{4}$ (6.7 W.u.) in $^{70}$Cr 
and 95 fm$^{4}$ (5.4 W.u.) in $^{72}$Fe. 
This 2qp transition strength becomes large because of the shallower Fermi level 
and the spatially extended structure of the quasiparticle wave function of the $\nu[431]1/2$ level.

At $N=42$, 
both of the $\nu[440]1/2$ and the $\nu[431]1/2$ levels are located far from the Fermi level, 
and 
2qp excitations of the $(\nu[301]1/2)^{2}$ and the $(\nu[422]5/2)^{2}$ have main contributions 
to the lowest $K^{\pi}=0^{+}$ state in $^{66}$Cr and $^{68}$Fe.  
These quadrupole transition matrix elements 
have opposite signs, but these transition strengths are less than 1 W.u. 

In Fig.~\ref{Cr_Fe_strength}, we show also the strengths for the 
monopole- and quadrupole-pair transitions. 
As in $^{34}$Mg, 
these low-lying $K^{\pi}=0^{+}$ modes have enhanced strengths for the quadrupole pair 
transition, whereas disappearing strengths for the monopole pair transition. 
This systematic calculation shows an importance of the dynamical pairing: 
Fluctuation of the pairing field, which is deformed as well as the mean field, 
is crucial for generating the collective $K^{\pi}=0^{+}$ modes. 

\section{\label{summary}Summary}
We have studied low-frequency $K^{\pi}=0^{+}$ modes in neutron-rich nuclei, taking account 
of the effects of nuclear deformation, pairing correlation 
and continuum coupling simultaneously. 
New type of this excitation mode is generated by coherent superposition of 
neutron 2qp excitations 
near the Fermi level whose wave functions have spatially extended structure. 
It is found that the dynamical pairing correlation, $i.e.$, the pairing vibration 
enhances its collectivity. 

In the spherical neutron-rich nuclei, the effect of the dynamical pairing 
has been investigated in detail in Refs.~\cite{mat01,mat05,paa03}. 
We have shown in this paper the importance of the dynamical pairing in neutron-rich deformed systems. 
In a deformed system where the up- and down-sloping orbitals exist near the Fermi level, 
one obtains the low-lying mode possessing extremely enhanced strengths both 
for the quadrupole p-h transition and for the quadrupole p-p (pair) 
transition induced by the pairing fluctuation. 

We found that the coupling between the pairing vibration 
and the neutron-skin vibration brings forth the soft $K^{\pi}=0^{+}$ mode 
in deformed Mg region with $N=22$. 
Furthermore, it was shown that emergence of low-lying $K^{\pi}=0^{+}$ modes 
is not restricted in the neutron-rich Mg isotopes. 
As an example, in neutron-rich Cr and Fe isotopes around $N=40$, 
we showed that the coherent coupling between the pairing vibration and 
the $\beta-$vibration of the neutron skin brings about the striking enhancement of the 
strengths for the quadrupole p-h and the quadrupole pair transition. 

\begin{acknowledgments}
The authors thank K.~Matsuyanagi for valuable comments and discussions. 
They also acknowledge N. Van Giai for useful discussions. 
One of the authors (K.Y) is supported by Research Fellowships of the 
Japan Society for the Promotion of Science for Young Scientists. 
The numerical calculations were performed on the NEC SX-8 supercomputers 
at Yukawa Institute for Theoretical Physics, Kyoto University and 
at Research Center for Nuclear Physics, Osaka University.

\end{acknowledgments}

\begin{appendix}
\section{Matrix elements for one-body operators}
Let us first consider matrix elements for particle-hole type one-body operators 
\begin{equation}
\langle ab|\hat{O}_{K}^{(\mathrm{uv})}|\mathrm{HFB}\rangle,
\end{equation}
where
\begin{equation}
\hat{O}^{(\mathrm{uv})}_{K}=\sum_{\sigma \sigma^{\prime}}\int d\boldsymbol{r} d\boldsymbol{r}^{\prime}
\delta_{\sigma,\sigma^{\prime}}\delta(\boldsymbol{r}-\boldsymbol{r}^{\prime})
O^{(\mathrm{uv})}_{K}(\boldsymbol{r})\hat{\psi}^{\dagger}(\boldsymbol{r}^{\prime}\sigma^{\prime})
\hat{\psi}(\boldsymbol{r}\sigma),
\end{equation}
and the HFB ground state and the 2qp excited states
\begin{subequations}
\begin{align}
\hat{\beta}_{i}|\mathrm{HFB}\rangle &= 0, \\
|ab\rangle &= \hat{\beta}^{\dagger}_{a}\hat{\beta}^{\dagger}_{b}|\mathrm{HFB}\rangle
\end{align}
\end{subequations}
are described by the quasiparticle operators. These operators are defined by the 
generalized Bogoliubov transformation 
\begin{subequations}
\begin{align}
\hat{\psi}^{\dagger}(\boldsymbol{r}\sigma)&=
\sum_{k}\varphi_{1,k}(\boldsymbol{r}\bar{\sigma})\hat{\beta}^{\dagger}_{k}
+\varphi^{*}_{2,k}(\boldsymbol{r}\sigma)\hat{\beta}_{k}, \\
\hat{\psi}(\boldsymbol{r}\sigma)&= 
\sum_{k}\varphi^{*}_{1,k}(\boldsymbol{r}\bar{\sigma})\hat{\beta}_{k}
+\varphi_{2,k}(\boldsymbol{r}\sigma)\hat{\beta}^{\dagger}_{k},
\end{align}
\end{subequations}
where 
\begin{equation}
\varphi_{k}(\boldsymbol{r}\bar{\sigma})=-2\sigma \varphi_{k}(\boldsymbol{r} -\!\sigma).
\end{equation}
In Appendices, we omit the subscript $\tau$ for simplicity.

The 2qp transition matrix elements are calculated as
\begin{align}
&\langle ab|\hat{O}_{K}^{(\mathrm{uv})}|\mathrm{HFB}\rangle \notag \\
=&
\langle \mathrm{HFB}|\hat{\beta}_{b}\hat{\beta}_{a} \hat{O}_{K}^{(\mathrm{uv})}|\mathrm{HFB} \rangle \notag \\
=&\sum_{kk^{\prime}}\sum_{\sigma \sigma^{\prime}}\int d\boldsymbol{r}d\boldsymbol{r}^{\prime}
\delta_{\sigma,\sigma^{\prime}}\delta(\boldsymbol{r}-\boldsymbol{r}^{\prime})
O^{\mathrm{(uv)}}_{K}(\boldsymbol{r}) \notag \\
&\times \varphi_{1,k}(\boldsymbol{r}^{\prime}\bar{\sigma}^{\prime})
\varphi_{2,k^{\prime}}(\boldsymbol{r}\sigma)
\langle \mathrm{HFB}|\hat{\beta}_{b}\hat{\beta}_{a}\hat{\beta}^{\dagger}_{k}\hat{\beta}^{\dagger}_{k^{\prime}}|\mathrm{HFB}\rangle \notag \\
=& \int d\boldsymbol{r}O^{(\mathrm{uv})}_{K}(\boldsymbol{r})
\{ -\varphi_{1,a}(\boldsymbol{r}\downarrow)\varphi_{2,b}(\boldsymbol{r}\uparrow)
+\varphi_{1,a}(\boldsymbol{r}\uparrow)\varphi_{2,b}(\boldsymbol{r}\downarrow) \notag \\
& +\varphi_{1,b}(\boldsymbol{r}\downarrow)\varphi_{2,a}(\boldsymbol{r}\uparrow)
-\varphi_{1,b}(\boldsymbol{r}\uparrow)\varphi_{2,a}(\boldsymbol{r}\downarrow) \} \notag \\
\equiv& O_{K,ab}^{(\mathrm{uv})}
\end{align}
using the quasiparticle wave functions, 
and we employed the Wick's theorem. In the cylindrical coordinate representation, we can rewrite as
\begin{align}
&\langle ab|\hat{O}_{K}^{(\mathrm{uv})}|\mathrm{HFB}\rangle = \notag \\
&2\pi\delta_{K,\Omega_{a}+\Omega_{b}}\int\!\! \rho d\rho dz
O^{(\mathrm{uv})}_{K}(\rho,z) \notag \\
&\times\{ \varphi_{1,a}(\rho,z,\uparrow)\varphi_{2,b}(\rho,z,\downarrow)
- \varphi_{1,a}(\rho,z,\downarrow)\varphi_{2,b}(\rho,z,\uparrow)  \notag \\
&\hspace{0.2cm}-\varphi_{1,b}(\rho,z,\uparrow)\varphi_{2,a}(\rho,z,\downarrow) 
+ \varphi_{1,b}(\rho,z,\downarrow)\varphi_{2,a}(\rho,z,\uparrow) \}, \label{matrix_element_ph}
\end{align}
where 
\begin{equation}
O^{(\mathrm{uv})}_{K}(\rho,z)=O^{(\mathrm{uv})}_{K}(\boldsymbol{r})e^{iK\phi}. 
\end{equation}

Next we consider the pair creation operators consisting of nucleons with opposite direction of spins
\begin{equation}
\hat{O}^{(\mathrm{uu})}_{K}=\int d\boldsymbol{r} d\boldsymbol{r}^{\prime}
\delta(\boldsymbol{r}-\boldsymbol{r}^{\prime})
O^{(\mathrm{uu})}_{K}(\boldsymbol{r})
\hat{\psi}^{\dagger}(\boldsymbol{r}^{\prime}\uparrow)
\hat{\psi}^{\dagger}(\boldsymbol{r}\downarrow).
\end{equation}
The matrix elements read
\allowdisplaybreaks[4]
\begin{align}
&\langle ab|\hat{O}_{K}^{(\mathrm{uu})}|\mathrm{HFB}\rangle \notag \\
=&\langle \mathrm{HFB}|\hat{\beta}_{b}\hat{\beta}_{a} \hat{O}_{K}^{(\mathrm{uu})}|\mathrm{HFB} \rangle \notag \\
=&\sum_{kk^{\prime}} \int d\boldsymbol{r}d\boldsymbol{r}^{\prime}
\delta(\boldsymbol{r}-\boldsymbol{r}^{\prime})
O^{\mathrm{(uu)}}_{K}(\boldsymbol{r}) \notag \\
& \times \varphi_{1,k}(\boldsymbol{r}^{\prime}\bar{\uparrow})
\varphi_{1,k^{\prime}}(\boldsymbol{r}\bar{\downarrow})
\langle \mathrm{HFB}|\hat{\beta}_{b}\hat{\beta}_{a}\hat{\beta}^{\dagger}_{k}\hat{\beta}^{\dagger}_{k^{\prime}}|\mathrm{HFB}\rangle \notag \\
=&\int d\boldsymbol{r}O^{(\mathrm{uu})}_{K}(\boldsymbol{r})
\{ -\varphi_{1,a}(\boldsymbol{r}\downarrow)\varphi_{1,b}(\boldsymbol{r}\uparrow)
+ \varphi_{1,b}(\boldsymbol{r}\downarrow)\varphi_{1,a}(\boldsymbol{r}\uparrow)\} \notag \\
=&2\pi\delta_{K,\Omega_{a}+\Omega_{b}}\int \rho d\rho dz
O^{(\mathrm{uu})}_{K}(\rho,z) \notag \\
&\times\{ \varphi_{1,a}(\rho,z,\uparrow)\varphi_{1,b}(\rho,z,\downarrow)
- \varphi_{1,a}(\rho,z,\downarrow)\varphi_{1,b}(\rho,z,\uparrow) \}, \label{matrix_element_pp}
\end{align}
where
\begin{equation}
O^{(\mathrm{uu})}_{K}(\rho,z)=O^{(\mathrm{uu})}_{K}(\boldsymbol{r})e^{iK\phi}.
\end{equation}

\section{QRPA transition matrix elements}
In terms of the nucleon annihilation and creation operators 
in the coordinate representation,  
the quadrupole operator is represented as
\begin{equation} 
\hat{Q}_{2K}
=\sum_{\sigma}\int d\boldsymbol{r}r^{2}Y_{2,-K}(\hat{r})
\hat{\psi}^{\dagger}(\boldsymbol{r}\sigma)
\hat{\psi}(\boldsymbol{r}\sigma).
\end{equation}
The intrinsic matrix elements 
$\langle \lambda|\hat{Q}_{2K}|0 \rangle$ 
of the quadrupole operator between the excited state $|\lambda \rangle$ 
and the ground state $|0\rangle$ are given by
\begin{equation}
\langle \lambda|\hat{Q}_{2K}|0 \rangle=\sum_{ab}
Q_{2K,ab}^{(\mathrm{uv})} 
(f_{ab}^{\lambda}+g_{ab}^{\lambda})
=\sum_{ab}M_{2K,ab}^{(\mathrm{uv})}, \label{QRPA_matrix}
\end{equation}
where 
\begin{equation}
Q_{2K,ab}^{(\mathrm{uv})}
=\langle ab|\hat{Q}_{2K}|\mathrm{HFB}\rangle
\end{equation} 
calculated by using the quasiparticle wave-functions as (\ref{matrix_element_ph}). 

We can furthermore calculate the intrinsic matrix element of the pair creation operator as 
\begin{subequations}
\begin{align}
\langle \lambda | \hat{P}_{0}^{\dagger} |0\rangle &= \sum_{ab} 
(f^{\lambda}_{ab}P_{00,ab}^{(\mathrm{uu})} 
+ g^{\lambda}_{ab}P_{00,ab}^{(\mathrm{vv})})
=\sum_{ab}M_{00,ab}^{(\mathrm{add})}, \\
\langle \lambda | \hat{P}_{2K}^{\dagger} |0\rangle &= \sum_{ab} 
(f^{\lambda}_{ab}P_{2K,ab}^{(\mathrm{uu})} 
+ g^{\lambda}_{ab}P_{2K,ab}^{(\mathrm{vv})})
=\sum_{ab}M^{(\mathrm{add})}_{2K,ab}, 
\end{align}
\end{subequations}
where the monopole- and quadrupole-pair creation operators are defined
by Eq.(\ref{eq:pair_transition}). 
The matrix element of 2qp excitation $P^{(\mathrm{uu})}_{00,ab}$, 
$P^{(\mathrm{vv})}_{00,ab}$, $P^{(\mathrm{uu})}_{2K,ab}$ and 
$P^{(\mathrm{vv})}_{2K,ab}$ 
are given by (\ref{matrix_element_pp}) 
and by the similar expression for $\hat{O}_{K}^{(\mathrm{vv})}$.

\section{Calculation of the moment of inertia}
We estimate moments of inertia using the spurious solution of the RPA equation~\cite{RS} 
\begin{subequations}
\begin{align}
\begin{pmatrix}
A & B \\ B^{*} & A^{*} 
\end{pmatrix}_{\alpha\beta\gamma\delta}
\begin{pmatrix}
J_{x} \\ -J_{x}^{*}
\end{pmatrix}_{\gamma\delta}
&=0,
\\
\begin{pmatrix}
A & B \\ B^{*} & A^{*} 
\end{pmatrix}_{\alpha\beta\gamma\delta}
\begin{pmatrix}
\Theta \\ -\Theta^{*}
\end{pmatrix}_{\gamma\delta}
&=\dfrac{\hbar}{i} \dfrac{1}{\mathcal{J}_{\mathrm{TV}}}
\begin{pmatrix}
J_{x} \\ J_{x}^{*}
\end{pmatrix}_{\alpha\beta},
\end{align}
\end{subequations}
where $\hat{J}_{x}$ is the angular momentum operator consisting of the orbital part $\hat{l}_{x}$ 
and the spin part $\hat{s}_{x}$, 
and its conjugate operator $\hat{\Theta}$.

The Thouless-Valatin moment of inertia $\mathcal{J}_{\mathrm{TV}}$ is 
determined thorough the orthonormal condition
\begin{equation}
\begin{pmatrix}
J_{x}^{*} & J_{x}  
\end{pmatrix}_{\alpha\beta}
\begin{pmatrix}
\Theta \\ -\Theta^{*}  
\end{pmatrix}_{\gamma\delta}
=\dfrac{\hbar}{i}\delta_{\alpha\beta, \gamma\delta},
\end{equation}
as
\begin{equation}
{\cal J}_{\mathrm{TV}}=2\hbar^{2}\sum_{\alpha\beta \gamma\delta}
(J_{x})^{*}_{\alpha\beta}(A-B)^{-1}_{\alpha\beta \gamma\delta}
(J_{x})_{\gamma\delta}.
\end{equation}
Turning off the residual interaction, we obtain the expression for 
the Inglis-Belyaev moment of inertia $\mathcal{J}_{\mathrm{Belyaev}}$
\begin{equation}
\mathcal{J}_{\mathrm{Belyaev}}=2\hbar^{2} 
\sum_{\alpha\beta}\dfrac{|J_{x}|^{2}_{\alpha\beta}}{E_{\alpha}+E_{\beta}}.
\end{equation} 

\end{appendix}

\end{document}